\documentclass[aps,multicol,epsfig,twocolumn,eqsecnum]{revtex4}
\newcommand\be{\begin{eqnarray}}
\newcommand\ee{\end{eqnarray}}
\newcommand\ba{\begin{array}}
\newcommand\ea{\end{array}}
\def\r{\rangle}
\def\l{\langle}

\def\T{{\rm Tr}}

\def\cI{{\cal I}}

\def\cE{{\cal E}}

\def\openone{{\it I}}

\usepackage{graphicx}
\begin{document}
\title{Concurrence {\it vs} purity: Influence of local channels on Bell states of two qubits }
\author{M\'ario Ziman$^{1,2,3}$ and Vladim\'\i r Bu\v zek$^{1,3,4}$}
\address{
$^1$Research Center for Quantum Information, Slovak Academy of Sciences,
D\'ubravsk\'a cesta 9, 84511 Bratislava, Slovakia
\\
$^2$Faculty of Informatics, Masaryk University, Botanick\'a 68a, 60200
Brno, Czech Republic\\
$^3$Quniverse, L\'\i\v s\v cie \'udolie 116, 841 04 Bratislava, Slovakia\\
$^4$Abteilung f\"{u}r Quantenphysik, Universit\"at Ulm, 89069 Ulm,
Germany }
\begin{abstract}
We analyze how a maximally entangled state of two-qubits (e.g., the singlet $\psi_s$) is affected
by  action of local channels described by  completely positive maps $\cE$ . We analyze
the concurrence and the purity of states $\varrho_\cE=\cE\otimes\cI[\psi_s]$.
Using the concurrence-{\it vs}-purity phase diagram we characterize local channels $\cE$ by their action
on the singlet state $\psi_s$.
We specify a region of the concurrence-{\it vs.}-purity
diagram that is achievable from the singlet state via the action of unital channels.
We shown that
 even most general (including non-unital) local channels acting just on a single qubit
of the original singlet state cannot generate the maximally entangled mixed states (MEMS).
We study in detail various time evolutions of the original singlet state induced by
local Markovian semigroups. We show that the decoherence process is represented
in the concurrence-{\it vs.}-purity diagram
by a line that forms the lower bound of the achievable region for unital maps.
On the other hand, the depolarization process is represented by a line that forms the upper
bound of the region of maps induced by unital maps.
\end{abstract}

 \maketitle

%%%%%%%%%%%%%%%%%%%%%%%%%%%%%%%%%%%%%%%%%%%%%%%%%%%%%%%%%%%%%%%%%%%%%%%%%%%%%
%\begin{multicols}{2}
%%%%%%%%%%%%%%%%%%%%%%%%%%%%%%%%%%%%%%%%%%%%%%%%%%%%%%%%%%%%%%%%%%%%%%%%%%%%%
\section{Introduction}
%%%%%%%%%%%%%%%%%%%%%%%%%%%%%%%%%%%%%%%%%%%%%%%%%%%%%%%%%%%%%%%%%%%%%%%%%%%%%

From two well-established properties of the entanglement, namely, from the fact that
{\it i) interactions create entanglement} and from the fact
{\it ii) entanglement cannot be shared freely} (monogamy
\cite{ckw,koashi,yu,osborne}),
we can conclude that any non-unitary evolution of a single qubit that is entangled with another qubit is
accompanied with a deterioration of the original
entanglement between these two qubits.

The aim of this paper is to address the question how local
actions (channels) affect properties of quantum states of bi-partite systems. In particular,
we will analyze in detail how the entanglement and the purity of a two-qubit
system that has been originally prepared in a maximally entangled Bell state depend on the action
of a single-qubit channel, i.e. we assume that one of the qubits of the original Bell pair is
affected by an environment.

We have a two-fold {\bf task} in front of us:
First, we will analyze how local channels affect the entanglement and the purity of the
original Bell state. Second, we we study
how the time evolution (i.e., a one parametric
subset $\cE_t$ of the set of all completely-positive maps) can be represented as a one-parametric curve in the
concurrence {\it vs.} purity ``phase'' diagram.
We will focus our attention on Markovian evolutions, i.e. those one-parametric subsets
of channels,
for which the semigroup property $\cE_t\cE_s=\cE_{t+s}$ holds.
We will analyze in detail physical processes such
as decoherence, decay, quantum homogenization, etc. in terms of
the concurrence {\it vs.} purity phase diagram.

Let us first define those quantities that we shall use through  the
paper. The purity (or equivalently the ``mixedness'') of a state
that is described by the density operator $\varrho$
will be quantified by the function
\be
P(\varrho)=\T[\varrho^2]\, ,
\ee
which equals to unity for pure states and achieves its minimum for
maximally mixed state, i.e.
for the total mixture $\varrho=\frac{1}{d}\openone$, the purity achieves the minimal value
that is equal to $1/d$.

The entanglement between two quantum systems described by a density operator $\varrho_{AB}\equiv\varrho$
will be quantified by the function called the {\it tangle}
\be
\label{tangle}
\tau(\varrho)=\min_{\varrho=\sum_k q_k \psi_k}\sum_k q_k S_2(\psi_k)\;,
\ee
where $\psi_k$ denotes the projection onto a pure state $|\psi_k\r$.
The minimum in Eq.~(\ref{tangle})
is taken over all pure-state decompositions of the state
$\varrho$ while the function $S_2(\psi_k)=2[1-P(\T_B\psi_k)]$ is the so-called {\it linear
  entropy}. The quantity $C(\varrho)=\sqrt{\tau(\varrho)}$
(the square root of the tangle) is known in the literature as the {\it
  concurrence}. Wootters \cite{wootters} has derived a simple analytic
formula for the concurrence of two qubits in a state $\varrho$
\be
C(\varrho)=2\max\{\mu_j\}-\sum_j \mu_j\, ,
\ee
where $\mu_j$ are square roots of eigenvalues of the matrix
$R=\varrho(\sigma_y\otimes\sigma_y)\varrho^*(\sigma_y\otimes\sigma_y)$
and $\varrho^*$ denotes the complex conjugation of the original two-qubit density operator $\varrho$. From these definitions
it is obvious that the entanglement and the purity are
closely related quantities and that for a given two-qubit state $\varrho$ they cannot take arbitrary values.

One of the questions one can ask at this point is:
Which two-qubit states are maximally entangled
providing that their purity is fixed, and {\it vice versa}?
This problem has been addressed in several earlier papers
\cite{ishizaka,verstraete,wei,adesso,batle,zyckowski}.
In particular, Ishizaka and Hiroshima \cite{ishizaka}
have introduced the so-called {\it maximally
entangled mixed states} (MEMS). These are the states that for a given value
of the purity achieve the maximal entanglement.
In Ref.~\cite{verstraete} a slightly more general problem
has been solved. The authors have shown which unitary transformation
has to be applied on a given state $\varrho$ in order to maximize the
entanglement.
In other words: which state maximizes the entanglement for a given
spectrum of the density operator (i.e., for a given value of the purity).
After these introductory papers have appeared many different aspects of
the relation between the entanglement and the mixedness have been
analyzed. In addition, the entanglement-based ``ordering'' (parametrization) of the state space of two-qubits,
originally introduced by Eisert and Plenio in \cite{eisert}, has been investigated in detail.
It has been shown, that different entanglement measures define different ordering of
states \cite{grudka}. In fact, this feature is not only characteristic for entanglement measures, but also for different measures
of mixedness. This means that the choice of the measures affects the
final entanglement-purity picture of the state space. In Ref.~\cite{wei}
the analysis of the entanglement-purity dependence for various
measures has been analyzed.

In this paper we will study the entanglement-purity relation
from a perspective of
{\it local} operations. In what follows we will assume that the
initial state of two qubits is a maximally entangled pure state, i.e. the two qubits are prepared in a Bell state.
Without the loss of generality we can consider that the two qubits are prepared in the
singlet state
\be
\label{singlet}
\psi_s=\frac{1}{4}(\openone\otimes\openone
-\sigma_x\otimes\sigma_x-\sigma_y\otimes\sigma_y-\sigma_z\otimes\sigma_z)\; .
\ee
This state is transformed under the action of a completely positive trace-preserving linear map
$\cE$
\cite{nielsen,presskill} that describes the most general quantum
process into the state
\be
\varrho_\cE=\cE\otimes\cI[\psi_s]\; .
\ee
 In general, any local action $\cE$ (except for unitary operations)
{\it decreases} the purity of the singlet state.
Our aim is to find how much the entanglement is changed under the action of the map
$\cE\otimes\cI$.

In Sections II-IV we will analyze in detail the action
of unital channels (i.e. those channels that do not affect the
total mixture, i.e. $\cE[I]=I$). The unital channels are defined in Sec.~II.
In Sec.~III we present a geometrical representation of the space of all unital maps.
In Sec.~IV we introduce the concurrence-{\it vs}-purity phase diagram and we determine
the region that is covered by the states $\varrho_\cE=\cE\otimes\cI[\psi_s]$ that are
obtained via the action of local unital maps on the singlet state of two qubits.

The Section
V is devoted to investigation of the action of  non-unital channels.
Finally, in Section VI we will
discuss the time evolution in the concurrence-purity (C-P) phase diagram for
specific quantum processes. In Conclusion we will summarize the main results
and discuss some open problems.

%%%%%%%%%%%%%%%%%%%%%%%%%%%%%%%%%%%%%%%%%%%%%%%%%%%%%%%%%%%%%%%%%%%%%%%%%%%%%
\section{Unital channels}
%%%%%%%%%%%%%%%%%%%%%%%%%%%%%%%%%%%%%%%%%%%%%%%%%%%%%%%%%%%%%%%%%%%%%%%%%%%%%
Let us consider firstly the unital channels. Thanks to a
seminal work of Ruskai et al. \cite{ruskai}
the investigation (parametrization) of single-qubit channels can be
significantly simplified. A single qubit channel
$\cE$ (not only unital ones) can be
written as a sequence of two unitary rotations and one specific
completely positive map $\Phi_\cE$ which belongs to a 6 parametric
family of maps. In particular, $\cE[\varrho]=U\Phi_\cE[V\varrho V^\dagger]
U^\dagger$. Due to the fact that unitary operations preserve essentially all
interesting properties of the original channel $\cE$, one can reduce
the analysis of  single-qubit channels into the investigation of  properties of the channel $\Phi_\cE$.
In other words, up to a unitary equivalence the original 15 parametric set
of single-qubit channels can be reduced into a 6 parametric set of channels
$\Phi_\cE$. This reduction significantly simplifies analysis of
single-qubit channels. In the Bloch sphere picture the general channel $\cE$
transforms the Bloch vector $\vec{r}$ in an {\it affine} way, i.e.
$\vec{r}\to\vec{r}^\prime=T\vec{r}+\vec{t}$. The corresponding map
$\Phi_\cE$ acts as follows
\be
\vec{r}\to\vec{r}^\prime=D\vec{r}+\vec{\tau}\; ,
\ee
where $D={\rm diag}\{\lambda_1,\lambda_2,\lambda_3\}$
is a diagonal matrix of singular values of the matrix $T$ and
$\vec{\tau} = R_U\vec{t}$ with $R_U$ being a three-dimensional rotation
associated with the unitary transformation $U$. The vector
$\vec{\tau}$ represents the shift of the total mixture, i.e. it
is associated with the non-unitality of the channel under consideration.

A set of unital channels $\Phi_\cE$ form a three-parametric family of completely positive (CP)
maps of the form $\Phi_\cE={\rm  diag}\{1,\lambda_1,\lambda_2,\lambda_3\}$
and the inequalities
\be
\nonumber
1+\lambda_x-\lambda_y-\lambda_z &\ge& 0 \, ;\\
\nonumber
1-\lambda_x+\lambda_y-\lambda_z &\ge& 0 \, ;\\
1-\lambda_x-\lambda_y+\lambda_z &\ge& 0 \, ;\\
\nonumber
1+\lambda_x+\lambda_y+\lambda_z &\ge& 0 \, ,
\ee
guarantee the complete positivity of these maps.

Our task is to evaluate the purity and the concurrence of states
$\Omega_\cE=\Phi_\cE\otimes\cI[\psi_s]$
(with $\psi_s=\frac{1}{4}(I\otimes I-\sigma_x\otimes\sigma_x
-\sigma_y\otimes\sigma_y-\sigma_z\otimes\sigma_z)$) as functions
of these three-parameters. The state $\Omega_\cE$ takes a simple form
$\Omega_\cE=\frac{1}{4}(I-\lambda_x\sigma_x\otimes\sigma_x-\lambda_y\sigma_y\otimes\sigma_y-\lambda_z\sigma_z\otimes\sigma_z)$.
The corresponding matrix of this state reads
\be
\label{omega}
\Omega_\cE=\left(
\begin{array}{cccc}
A & 0 & 0 & D \\
0 & B & C & 0 \\
0 & C & B & 0 \\
D & 0 & 0 & A \\
\end{array}\right)
\ee
with $A=\frac{1}{4}(1-\lambda_z)$, $B=\frac{1}{4}(1+\lambda_z)$,
$C=-\frac{1}{4}(\lambda_y+\lambda_x)$ and
$D=\frac{1}{4}(\lambda_y-\lambda_x)$.

In order to evaluate the purity of the state $\Omega_\cE$ we have to find eigenvalues of the matrix (\ref{omega}).
These eigenvalues are given by the expression $\kappa_{1,2}=A\pm D$ and
$\kappa_{3,4}=B\pm C$. Thus, for the purity of the state $\Omega_\cE$ we find the expression
\be
P(\Omega_\cE)=\T[\Omega_\cE^2]=\frac{1}{4}(1+\lambda_x^2+
\lambda_y^2+\lambda_z^2)\; .
\ee

In order to find the concurrence of the state $\Omega_\cE$ we have to
evaluate the eigenvalues of the matrix
\be
R=\Omega_\cE\sigma_y\otimes\sigma_y\Omega_\cE\sigma_y\otimes\sigma_y=\left(
\ba{cccc}
X & 0 & 0 & Y \\
0 & P & Q & 0 \\
0 & Q & P & 0 \\
Y & 0 & 0 & X \\
\ea\right)
\ee
with $X=A^2+D^2$, $Y=-2AD$, $P=C^2+B^2$, $Q=2CB$.
The square roots of these eigenvalues are $\mu_{1,2}=|B\pm C|$
and $\mu_{3,4}=|A\pm D|$. Since the eigenvalues of $\Omega_\cE$ are
positive, the square roots of eigenvalues of $R$ and the eigenvalues of
$\Omega_\cE$ coincide, i.e. the absolute values can be removed
and the concurrence is given by the formula
\be
\label{unit_conc}
C(\Omega_\cE)=\frac{1}{2}\max\left\{
\ba{c}
\lambda_x+\lambda_y+\lambda_z-1\\
\lambda_x-\lambda_y-\lambda_z-1\\
-\lambda_x+\lambda_y-\lambda_z-1\\
-\lambda_x-\lambda_y+\lambda_z-1
\ea\, , 0\right\}\; .
\ee

%%%%%%%%%%%%%%%%%%%%%%%%%%%%%%%%%%%%%%%%%%%%%%%%%%%%%%%%%%%%%%%%%%%%%%%%%%%%%
\section{Parametrization of local CP maps: Geometric picture}
%%%%%%%%%%%%%%%%%%%%%%%%%%%%%%%%%%%%%%%%%%%%%%%%%%%%%%%%%%%%%%%%%%%%%%%%%%%%%
We have derived explicit relations for purity and concurrence of a two-qubit
density operator $\Omega_\cE$ that is obtained via the action the unital channel $\cE$
on the singlet state (\ref{singlet}). Unfortunately,
the concurrence is not so easy to deal with. Let us
illustrate the whole situation in the space of parameters
$\lambda_x,\lambda_y,\lambda_z$. It is a well known result of the
analysis of qubit channels that unital channels (specified by $\lambda$'s)
form a tetrahedron with unitary Pauli operators in its vertices.
The same geometrical picture holds for states of the form $\Omega_\cE$. The
extremal points of this tetrahedron are mutually orthogonal maximally entangled
states (the Bell basis). The convex combinations of these states form the
tetrahedron, i.e. a classical probability simplex.

The states of the same purity correspond to a sphere of the
radius proportional to a specific value of the purity centered at the point
$\lambda_x=\lambda_y=\lambda_z=0$. In particular,
$|\vec{\lambda}|=4P-1$. There are only four pure
states represented by the intersection of the tetrahedron
with the sphere of the radius $|\vec{\lambda}|=3$, i.e. the sphere in which the
tetrahedron is embedded. These points are exactly
the four maximally entangled states forming the Bell basis.

The equally entangled states specify planes (that form a polytope inside
the tetrahedron). Because of the discrete symmetry represented by
four unitary transformations (rotations) described by the operators
$I,\sigma_x,\sigma_y,\sigma_z$, the analysis of possible
values of the concurrence and the purity in terms of $\vec{\lambda}$
can be focused onto only two cases: i) all $\lambda_j$ are positive,
or ii) all $\lambda_j$ are negative. The symmetry relating these
two options is the space inversion ($\vec{\lambda}\to\vec{-\lambda}$),
which is not physical (i.e., this inversion cannot be realized by a CP map).
In fact, the tetrahedron does not possess
such symmetry. The tetrahedron is symmetric under rotations
by an angle $\phi=\pi/2$ along each axis ($\sigma$ matrices).
Consequently, the analysis of the maximum
of the concurrence reduces to an investigation of two
cases (see Fig.~1): $\lambda_j\ge 0$ (for all $j$)
and $\lambda_j\le 0$ (for all $j$).

Let us start with the case of positive values of $\lambda$s. In this case the
maximum of the concurrence (\ref{unit_conc}) is achieved for
$C(\Omega_\cE)=\frac{1}{2}(\lambda_x+\lambda_y+\lambda_z-1)$. Providing that this number
is larger than zero, then states of the same concurrence $C$
form the plane $\lambda_x+\lambda_y+\lambda_z=d$ with $d=2C+1$.
These ``iso-concurrence'' planes intersecting
the tetrahedron (its positive octant) are given by normal vectors
$\vec{n}=(1,1,1)$, i.e. $\vec{n}\cdot\vec{\lambda}=2C+1$. The plane
with $C=0$ ($d=1$) form a boundary (face) of separable states in this
part of the tetrahedron. Due to the symmetry  mentioned above the same
picture holds for other four vertices of the tetrahedron,
i.e. tetrahedrons ``under'' four maximally entangled states (that correspond to vertices of the tetrahedron).

%%%%%%%%%%%%%%%%%%%%%%%%%%%%%%%%%%%
\begin{figure}
\begin{center}
\includegraphics[width=7cm]{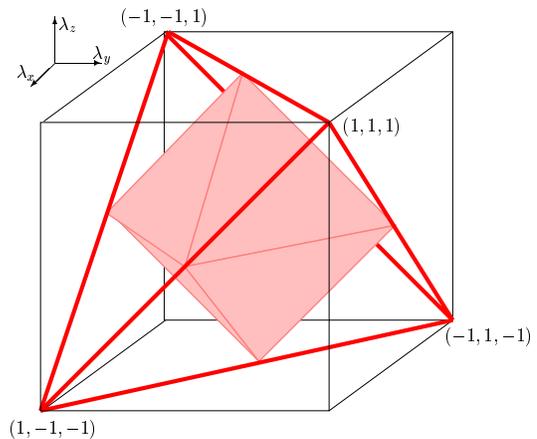}
\caption{
(Color online) This figure presents the space of unital channels $\Phi_\cE$ parameterized by
$\vec{\lambda}=(\lambda_1,\lambda_2,\lambda_3)$
(with $-1\le\lambda_j\le 1$).
Up to unitary rotations the set of states
$\Omega_\cE$ for that are obtained by the action of unital channels $\cE$ on the singlet state
(\ref{singlet}) form a
tetrahedron with vertices corresponding to maximally entangled states
that form the Bell basis. The octahedron inside the tetrahedron
represents a set of all separable states. The remaining four
tetrahedrons in each corner of the original tetrahedron
represent sets of entangled states.  The closer the point is
to the vertex the more
entangled the corresponding state is. The planes parallel to
faces of the separable
region contain states with the same degree of entanglement
(``iso-concurrence'' planes). States with
the same purity form spheres  (``iso-purity'' spheres) centered at the
center of the cube,
i.e. at the point $\vec{\lambda}=(0,0,0)$.
}
\end{center}
\label{tetrahedron}
\end{figure}
%%%%%%%%%%%%%%%%%%%%%%%%%%%%%%%%%%%%%

The negative region of the tetrahedron
($\lambda_j\le 0$) contains no entangled states, i.e. it consists of
only separable states. One can prove this by
analyzing all possibilities, or also by exploiting the geometrical
picture. The set of allowable (negative) $\vec{\lambda}$ is bounded
by the plane $1+\lambda_x+\lambda_y+\lambda_z=0$ that  potentially
contains entangled states in this ``negative region''. If not, then
due to a convexity of separable states the whole ``negative region''
is separable. Using the expressions for
$\lambda_x,\lambda_y,\lambda_z$
determined by the equation of the plane and by calculating the concurrence we
obtain $C=\frac{1}{2}\max\{-2,2\lambda_x,2\lambda_y,2\lambda_z,0\}$.
Because of $\lambda_j\le 0$ we obtain that the concurrence always equals to zero.

So far, we have shown that  entangled states belong to regions close to the vertices
of the tetrahedron of all possible states $\Omega_\cE$. Separable
states form the octahedron embedded in the tetrahedron of all states (see Fig.~1). Our next task is to evaluate the
concurrence and the purity for all entangled states. In particular,
we are interested in the shape of the region formed by the
states $\Omega_\cE$ in the C-P phase diagram.

%%%%%%%%%%%%%%%%%%%%%%%%%%%%%%%%%%%%%%%%%%%%%%%%%%%%%%%%%%%%%%%%%%%%%%%%%%%%%
\section{Concurrence {\it vs} purity diagram}
%%%%%%%%%%%%%%%%%%%%%%%%%%%%%%%%%%%%%%%%%%%%%%%%%%%%%%%%%%%%%%%%%%%%%%%%%%%%%
One possibility how to characterize bi-partite states is to specify their
purity and the value of their entanglement \cite{ishizaka}. Such classification contains
highly non-trivial information about the state itself. One can choose
different measures for both the purity and the entanglement. The
resulting characterization strongly depends on the particular choice
of measures of entanglement and the purity \cite{verstraete,wei}.
In what follows we will use two most common measures:
the mixedness for the purity and the concurrence for the entanglement.

As we have already mentioned the boundaries of the C-P phase diagram has
been  analyzed and the
results are known. The states maximizing the concurrence for the
given value of purity are known \cite{ishizaka} as {\it maximally
  entangled mixed states} (MEMS).
This concept generalizes the notion of Bell states, i.e.
maximally entangled pure states. MEMS have the following form (up to
local unitary transformations):
\be
\varrho_{\rm MEMS}=p |\phi_+\r\l\phi_+|+(1-p)|01\r\l 01|
\ee
for $p\in[2/3,1]$ and
\be
\varrho_{\rm MEMS}&=&p|\phi_+\r\l\phi_+|+\frac{1}{3}|01\r\l 01|\\
& + &
(\frac{1}{3}-p/2)(|00\r\l 00|+|11\r\l 11|)
\nonumber
\ee
for $p\in[0,2/3]$ and $|\phi_+\r=\frac{1}{\sqrt{2}}(|00\r+|11\r)$.
These states specify the region in the C-P phase diagram
which is physical, i.e. any point in this part of the C-P phase diagram corresponds to a state of quantum-mechanical system
(see Fig.~2).
Our aim is to analyze this picture
for the states of the form $\Omega_\cE$, i.e. those states that are obtained
by the action of local unital operations on the maximally entangled state $\psi_s$.
In particular, our task is to find the maximum/minimum value of the
concurrence for the given value of the purity.

As we have already shown states of equal purity correspond to the sphere (parameterized by
$\lambda_x^2+\lambda_y^2+\lambda_z^2=4P-1$) in the tetrahedron
of all states $\Omega_\cE$ (see Fig.~1). States with the same amount of entanglement
determine a plane $\lambda_x+\lambda_y+\lambda_z=2C+1$.
We have argued that it is sufficient to consider only the positive
region ($\lambda_j\ge 0$) of the tetrahedron. Given the iso-purity sphere
the iso-entanglement plane with the maximal value of entanglement is the one that
``intersects'' the sphere only in a single point. One can exploit the
geometric picture to see that this
point (state) fulfills the condition
$\lambda_x=\lambda_y=\lambda_z\equiv\lambda$. After we insert this condition
into the equations for the purity and the concurrence we obtain
\be
\left.
\ba{c}
3\lambda^2=4P-1 \\
3\lambda=2C+1
\ea\right\}\
\Rightarrow \ C_{\max}=\frac{1}{2}(\sqrt{3(4P-1)}-1)\; .
\ee
These states form the upper bound of the available region
in the C-P phase diagram (see Fig.~2). In particular, this bound contains
(is formed by)  Werner states, i.e.
$\varrho_W= q \psi_s + (1-q)\frac{1}{4}I$ and $C_{\max}=C_W$.

%%%%%%%%%%%%%%%%%%%%%%%%%%%%%%%%%%%
\begin{figure}
\label{cepe_diagram}
\begin{center}
\includegraphics[width=7cm]{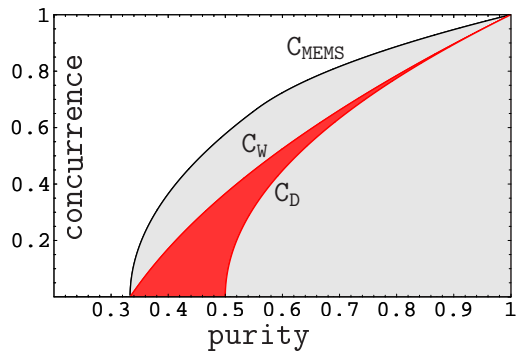}
\caption{(Color online)
The concurrence vs. purity phase diagram of
two qubits. The region of states $\Omega_\cE=\cE\otimes\cI[\psi_s]$
(region colored in red in the C-P phase diagram) that are obtained from
the maximally
entangled two-qubit states by the action of unital channels
$\cE$ is bounded by two lines corresponding
to $C_{W}$ and $C_{D}$ for a given value of the purity.
The upper bound of the region of physically realizable states (shaded sector of the C-P diagram)
corresponds to maximally entangled mixed
states (MEMS). We note that the MEMS states cannot be achieved
from Bell state by applying local channels on a {\it single} sub-system only.
}
\end{center}
\end{figure}
%%%%%%%%%%%%%%%%%%%%%%%%%%%%%%%%%%%%%
The next step is to analyze the minimum of the concurrence for a given value of purity.
In general this minimum is trivial, because there always exist
separable states for a given value of purity. However, in our case we investigate
states that are generated by local unital channels from the maximally entangled Bell states.
In this case a nontrivial lower bound exists. Exploiting the geometry of the states
$\Omega_\cE$  we can conclude that this minimum is zero
for all purity spheres for which the intersection with the tetrahedron
contains the plane $\lambda_1+\lambda_2+\lambda_3=1$ ($C=0$), i.e. the
boundary of the separable states. The states of the maximal purity
belonging to this plane (i.e., the maximally mixed separable states) are
the points with two same components and the remaining one
equals to unity. Without the loss of generality let us consider the
case $\lambda_z=1$ and $\lambda_x=\lambda_y\equiv \lambda$.
The equation of the plane implies that $\lambda=0$ and consequently, the
purity $P=\frac{1}{4}(1+|\vec{\lambda}|^2)=\frac{1}{2}$. That is,
for the states with the purity larger than 1/2 there are no separable
states $\Omega_\cE$.

The question on the minimum value of entanglement for a
given value of the purity is equivalent to the question about the maximum of the purity
for a given value of the concurrence. Using the same arguments as before
we find that the state of the maximal purity
satisfy the same conditions, i.e. $\lambda_z=1$
and $\lambda_x=\lambda_y\equiv\lambda$. Consequently, the equation of plane
$\lambda_x+\lambda_y+\lambda_z=1+2C$ implies $\lambda=C$.
As a result we obtain that $P=\frac{1}{2}(1+C^2)$. Inverting this
formula we obtain the functional dependence of the concurrence as a function of the purity
that specifies the lower bound of the allowable region in C-P phase diagram
for states $\Omega_\cE$
\be
C_{\min}=\sqrt{2P-1}\, .
\ee
In conclusion, states induced by local unital channels from the initial maximally entangled state
are represented in the concurrence {\it vs} purity diagram by points in the region that is bounded from
above by the ``line'' $C_{\max}=C_W$ and from
the below by the line $C_{\min}$ (see the red region in Fig.~2).
 The whole region of physically
relevant states is determined by the line $C_{\rm MEMS}$ corresponding to the MEMS.
Unital channels do not allow us to achieve states that are within the lines $C_{W}$ and $C_{\rm MEMS}$.
Simultaneously, we can conclude that the states $\Omega_\cE$ with the purity
larger than 1/2 remain entangled (this property is given by the lower
bound $C_{\min}=C_D$).

%%%%%%%%%%%%%%%%%%%%%%%%%%%%%%%%%%%%%%%%%%%%%%%%%%%%%%%%%%%%%%%%%%%%%%%%%%%%%
\section{Non-unital channels}
%%%%%%%%%%%%%%%%%%%%%%%%%%%%%%%%%%%%%%%%%%%%%%%%%%%%%%%%%%%%%%%%%%%%%%%%%%%%%
In the previous section we have shown that unital channels (up to unitary rotations) are characterized
by three parameters.
The non-unital channels are characterized
by six parameters. The triple $\vec{\tau}=(\tau_x,\tau_y,\tau_z)$ describes
how the total mixture is transformed under the action of non-unital channels.
In particular, under the action of a general non-unital map $\cE$, the
maximally entangled state (e.g., the singlet) is transformed into the state
\be
\Omega_\cE=\frac{1}{4}\left[
(\openone+\vec{\tau}\cdot\vec{\sigma})\otimes\openone
-\vec{\lambda}\cdot(\vec{\sigma}\otimes\vec{\sigma})
\right]\; ,
\ee
where $\vec{\lambda}\cdot(\vec{\sigma}\otimes\vec{\sigma})
=\lambda_x\sigma_x\otimes\sigma_x
+\lambda_y\sigma_y\otimes\sigma_y
+\lambda_z\sigma_z\otimes\sigma_z$.
In other words
\be
\Omega_\cE=\left(
\ba{cccc}
A+\tau & F & 0 & D \\
F^* & B-\tau & C & 0 \\
0 & C & B+\tau & F \\
D & 0 & F^* & A-\tau \\
\ea\right)\; ,
\ee
where $A,B,C,D$ are defined as before and $F=(\tau_x-i\tau_y)/4$,
$\tau=\tau_z/4$.
First we address the question whether by using  non-unital channels the
upper bound on concurrence can be increased, in particular, whether
$\Omega_\cE$ can be a MEMS state (i.e. the state on the $C_{\rm MEMS}$ line). In order to answer this question we  use the
identity $\T_A\Omega_\cE=\T_A \psi_s=\frac{1}{2}\openone$ which holds for any {\it local} action $\cE\otimes\cI$.
For MEMS states we find that
$\T_A\varrho_{\rm MEMS}\ne \frac{1}{2}\openone$
and $\T_B\varrho_{\rm MEMS}\ne \frac{1}{2}\openone$ as well, i.e.
$\Omega_\cE$ cannot be a MEMS state.
Therefore, we conclude that the MEMS states cannot be achieved from the maximally entangled state via local operations, i.e. MEMS
are not the states of the form $\Omega_\cE=\Phi_\cE\otimes\cI[\psi_s]$.

However, the action of non-unital channels on maximally entangled states can be different then the action
of unital channels. That is, the states $\Omega_\cE=\Phi_\cE\otimes\cI[\psi_s]$
that are obtained by the action of non-unital channels might lie in the region of the concurrence {\it vs} purity phase diagram that
are bounded by the lines $C_{\rm MEMS}$ and $C_{W}$.
The purity of states $\Omega_\cE$ can be calculated by finding the
trace of $\Omega_\cE^2$. A direct calculation gives that
$P=\frac{1}{4}(1+|\vec{\lambda}|^2+|\vec{\tau}|^2)$. Thus in the picture
of $\vec{\lambda}$ parameters (for a fixed vector $\vec{\tau}$) the
states of the same purity belong to the sphere
$|\vec{\lambda}|^2=4P-1-|\vec{\tau}|^2$. Note that for the fixed $\vec{\tau}$
the set of all possible $\vec{\lambda}$ do not form the tetrahedron
anymore.

Let us consider a specific case $\vec{\tau}=(0,0,\tau_z)$, i.e. $F=0$. In this
case the eigenvalues of $\Omega_\cE$ read:
\be
\nonumber
\mu_1&=&\frac{1}{4}(1-\lambda_z+\sqrt{(\lambda_x-\lambda_y)^2+\tau_z^2})\; ;\\
\nonumber
\mu_2&=&\frac{1}{4}(1-\lambda_z-\sqrt{(\lambda_x-\lambda_y)^2+\tau_z^2})\; ;\\
\nonumber
\mu_3&=&\frac{1}{4}(1+\lambda_z+\sqrt{(\lambda_x+\lambda_y)^2+\tau_z^2})\; ;\\
\nonumber
\mu_4&=&\frac{1}{4}(1+\lambda_z-\sqrt{(\lambda_x+\lambda_y)^2+\tau_z^2})\; .
\ee
The positivity of these eigenvalues determines the set of all allowed
values of parameters $\vec{\lambda},\tau_z$. Fixing $\tau_z$ we obtain four
surfaces (setting $\mu_j=0$) in the three-dimensional space
of parameters $\vec{\lambda}$ that form  boundaries of the set of
all possible states $\Omega_\cE$. The identities $\mu_j=0$ can
be rewritten into the equations
\be
\nonumber
\lambda_z&=&1-\sqrt{(\lambda_x+\lambda_y)^2+\tau_z^2}\; ;\\
\lambda_z&=&-1+\sqrt{(\lambda_x+\lambda_y)^2+\tau_z^2}\; ,
\ee
that completely specify the shape of the set in the space of parameters
$\lambda_1,\lambda_2,\lambda_3$. For non-unital channels the corners of
tetrahedron are smoothed (``rounded'') depending on the value of the shift
$\vec{\tau}$.

To compute the concurrence $C$ of state $\Omega_\cE=\Phi_\cE\otimes\cI[\psi_s]$
we have to find eigenvalues of the matrix
\be
\nonumber
R=\Omega_\cE(\sigma_y\otimes\sigma_y)\Omega_\cE^*(\sigma_y\otimes\sigma_y)=
\left(
\ba{cccc}
\alpha & 0 & 0 & \delta_+\\
0 & \beta &  \gamma_- & 0\\
0 & \gamma_+ & \beta & 0\\
\delta_- & 0 & 0 & \alpha
\ea
\right)\; ,
\ee
where $\alpha=A^2-\tau^2+D^2$, $\beta=B^2-\tau^2+C^2$,
$\gamma_\pm=2C(B\pm \tau)$, and $\delta_\pm=2D(A\pm \tau)$ ($\tau=\tau_z/4$).
Square roots of the eigenvalues of the matrix $R$ read
\be
\{\sqrt{A^2-\tau^2}\pm |D|, \sqrt{B^2-\tau^2}\pm |C|\}\; ,
\ee
and for the concurrence we obtain the expression
\be
C=\max\{0, 2(|D|-\sqrt{B^2-\tau^2}), 2(|C|-\sqrt{A^2-\tau^2})\}\, .
\ee
Using the parameters $\vec{\lambda},\tau_z$ the concurrence can be rewritten in the form
\be
C=\frac{1}{2}\max\left\{\ba{c}
0,\\ |\lambda_y-\lambda_x|-\sqrt{(1+\lambda_z)^2-\tau_z^2},\\
|\lambda_x+\lambda_y|-\sqrt{(1-\lambda_z)^2-\tau_z^2}\\
\ea\right\}\, .
\ee
As we will show in the next Section the states $\Omega_\cE=\Phi_\cE\otimes\cI[\psi_s]$ that are generated
by non-unital channels can lie above the line $C_{\max}$ in the concurrence {\it vs} purity diagram.
In order to have more physical insight into the action of the local channels on maximally entangled states
let us consider one-parametric set of local maps that correspond to specific time evolutions of two-qubit systems.

%%%%%%%%%%%%%%%%%%%%%%%%%%%%%%%%%%%%%%%%%%%%%%%%%%%%%%%%%%%%%%%%%%%%%%%%%%%%%
%%%%%%%%%%%%%%%%%%%%%%%%%%%%%%%%%%%%%%%%%%%%%%%%%%%%%%%%%%%%%%%%%%%%%%%%%%%%%
\section{Evolution in C-P diagram}
%%%%%%%%%%%%%%%%%%%%%%%%%%%%%%%%%%%%%%%%%%%%%%%%%%%%%%%%%%%%%%%%%%%%%%%%%%%%%
In  this section we will analyze how the evolution
of a maximally entangled state under the action of a
local channel is reflected
in the C-P diagram. The case of unitary evolution is trivial:
Under the action of local unitary transformations neither the concurrence nor the purity are changed. Therefore,
the state $\Omega_t=(U_t\otimes\openone) \psi_s (U_{-t}\otimes I)$
is still a maximally entangled pure state.
In what follows we will analyze several models of non-unitary dynamics.
We will focus our attention on Markovian semigroup dynamics.
In particular, we will consider processes of
the decoherence, the depolarization and the homogenization.

\subsection{Decoherence}

The decoherence of a qubit is induced by the master equation \cite{ziman_deco}
$\dot{\varrho}=i[H,\varrho]+(T/2)[H,[H,\varrho]]$. The solution
of this equation form a semigroup of unital quantum channels
\be
\cE_t=\left(\ba{cccc}
1 & 0 & 0 & 0 \\
0 & e^{-t/T}\cos{\omega t} & e^{-t/T}\sin{\omega t} & 0 \\
0 & -e^{-t/T}\sin{\omega t} & e^{-t/T}\cos{\omega t} & 0 \\
0 & 0 & 0 & 1
\ea\right)\; .
\ee
The singular values of these channels are
\be
\nonumber
&\lambda_1(t)=\lambda_2(t)= e^{-t/T}\; ; &\\
&\lambda_3(t)= 1\; . &
\ee
Due to the evolution of one qubit the singlet is transformed
into a state $\Omega_t$ with the purity
\be
P_t=\frac{1}{4}(1+|\vec{\lambda}_t|^2)=\frac{1}{2}(1+e^{-2t/T})
\ee
and with the concurrence
\be
C_t=\frac{1}{2}(\lambda_1(t)+\lambda_2(t)+\lambda_3(t)-1)=e^{-t/T}\; .
\ee
Comparing these two equations we find that the purity and the concurrence
of the state $\Omega_t=\Phi_\cE\otimes\cI[\psi_s]$ induced by decoherence acting on one qubit are related
as
\be
P_t=\frac{1}{2}(1+C_t^2)\; ,
\ee
or equivalently
\be
C_t=\sqrt{2P_t-1} \; .
\ee
Since $P_t\in [1/2,1]$ we have obtained that the process of single-qubit decoherence in a C-P
diagram corresponds to the lower bound of the allowed region for
unital channels, i.e. $C_t:=C_{D}=C_{min}$.

\subsection{Depolarization}
The process of a single-qubit depolarization in a specific basis is represented by the semigroup \cite{nielsen}
\be
\cE_t=\left(\ba{cccc}
1 & 0 & 0 & 0 \\
0 & e^{-t/T} & 0 & 0 \\
0 & 0 & e^{-t/T} & 0 \\
0 & 0 & 0 & e^{-t/T}
\ea\right)\; .
\ee
In this case the states $\Omega_t$ are Werner states, i.e. this type
of dynamics in C-P diagram
is represented by the line $C_{W}$
for unital channels. In particular,
\be
\Omega_t=e^{-t/T}\psi_s +(1-e^{-t/T})\frac{1}{4}\openone \; .
\ee
Thus, we have found two processes that saturates the upper and lower
bound of the region in C-P diagram that is allowed for unital channels. In particular,
the decoherence saturates the lower bound while the  depolarization process
defines the upper bound.

\subsection{Homogenization}
The process of quantum homogenization is described by
the  semigroup of non-unital channels \cite{ziman_homo,ziman_osid}
\be
\label{homo_map}
\cE_t=\left(\ba{ccc}
1 & 0 \ \ \ 0 & 0 \\
\ba{c} 0 \\ 0 \ea & e^{-t/T_2} R_{\omega t}  & \ba{c} 0 \\ 0 \ea \\
w(1-e^{-t/T_1}) & 0 \ \ 0 & e^{-t/T_1}
\ea\right)\; ,
\ee
where
\be
R_{\omega t}=\left(
\ba{cc}
\cos{\omega t} & \sin{\omega t}\\
-\sin{\omega t} & \cos{\omega t}
\ea\right)
\ee
is a rotation matrix. This process describes an evolution that
transforms the whole Bloch sphere into a single point, i.e. a
generalization of an exponential decay. That is, quantum homogenization is described by
a contractive map with the fixed point that is the stationary state of the dynamics.
The parameters in the description of the map (\ref{homo_map}) have the following meaning:
$w$ is the purity of the final state, $T_1$ is the decay time,
$T_2$ is the decoherence time, and $\omega$ describes the unitary
part of the evolution. The singular values are similar to those in the
decoherence, i.e.
\be
\nonumber
&\lambda_1(t)=\lambda_2(t)=e^{-t/T_2}\; ;& \\
&\lambda_3(t)=e^{-t/T_1}\; . &
\ee
The homogenization belongs to a class of  non-unital
channels that we have analyzed in previous section. Therefore we can easily find expressions for the purity
\be
P_t=\frac{1}{4}\left[ 1+2e^{-2t/T_2}+e^{-2t/T_1}+w^2(1-e^{-t/T_1})^2\right]
\ee
and the concurrence
\be
C_t=\max\{0, e^{-t/T_2}-\frac{1}{2}(1-e^{-t/T_1})\sqrt{1-w^2}\}\;.
\ee
The resulting lines for all considered evolutions (decoherence,
depolarization and homogenization for the values $T_1/T_2=1/2$ and $w=1$)
are depicted in Fig.~3.

In a special case, when $w=0$ (final state is the total mixture),
the homogenization process is unital. In this case the purity and the
concurrence are
\be
&P_t=\frac{1}{4}(1+e^{-2t/T_1}+2e^{-2t/T_2})&\\
&C_t=e^{-t/T_2}+\frac{1}{2}(e^{-t/T_1}-1)&
\ee
In Fig.~5 we can see the corresponding line
for different fraction of decay and decoherence times, i.e. $T_1$ and $T_2$,
respectively. Interesting point is when these two rates coincides ($T_1=T_2$)
when the homogenization saturates the Werner states line (i.e. $C_{W}$).
In the limit of $T_1/T_2\to\infty$ the homogenization approaches
to the decoherence line (i.e. $C_{D}$).

For $w=1$ the homogenization describes the exponential
decay to a pure state. In such case
\be
&P_t=\frac{1}{2}(1+e^{-2t/T_1}+e^{-2t/T_2}-e^{-t/T_1})&\\
&C_t=e^{-t/T_2}&
\ee
In this case one can express the purity as a function of concurrence,
i.e. $P=\frac{1}{2}(1+C^2+C^{2T_2/T_1}-C^{T_2/T_1})$. In the special
case $T_2=2T_1$ (i.e. decoherence time is twice as fast as decay time)
the homogenization follows the line
\be
C=\sqrt[4]{2P-1}\; .
\ee
This line goes above the region determined by unital
channels (see Fig.~3).

In Figs.~4-6 we analyze various regimes of the homogenization process: We consider
processes with a fixed value of the ratio $T_1/T_2=1/2$,
but we change fixed points of the evolution
(Fig.~4). In Fig.~5 we consider the homogenization process with the fixed point being equal to the total mixture
(i.e. in this case the homogenization is a unital process) and we consider
different values of the ratio $T_1/T_2$. In Fig.~6 we consider the homogenization process with the fixed point equal to
a pure state (i.e. in this case the homogenization is a non-unital process) and we consider
different values of the ratio $T_1/T_2$.

%%%%%%%%%%%%%%%%%%%%%%%%%%%%%%%%%%%
\begin{figure}
\label{cepe_process}
\begin{center}
\includegraphics[width=7cm]{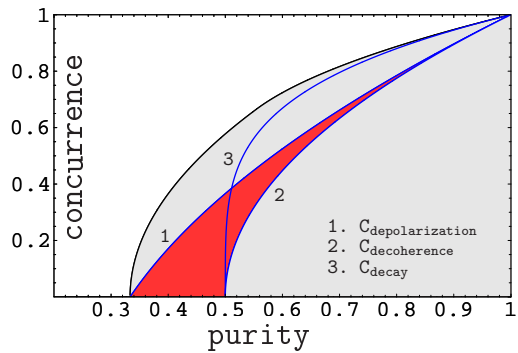}
\caption{(Color online)
Parametric plots of time evolution of the concurrence and the purity in
processes of the decoherence, the homogenization
and the depolarization. The lines are parameterized in such way that
the initial moment of the time evolution $t=0$ corresponds to the point
$C=P=1$ (upper right corner of the C-P diagram).
The decoherence line (2) represents the lower bound ($C_{D}$) of the
unital region, the
depolarization line (1) forms the upper bound ($C_{W}$) of the unital region.
The homogenization (a non-unital process) is characterized by the line (3)
that is outside the unital
region. In the present case the homogenization describes an exponential
decay, i.e. the fixed point of the evolution is the state $|0\rangle$.
}
\end{center}
\end{figure}
%%%%%%%%%%%%%%%%%%%%%%%%%%%%%%%%%%%%%

%%%%%%%%%%%%%%%%%%%%%%%%%%%%%%%%%%%
\begin{figure}
\label{cepe_homo}
\begin{center}
\includegraphics[width=7cm]{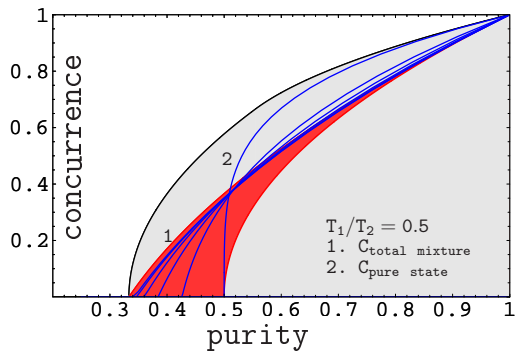}
\caption{(Color online)
We present evolutions of the concurrence and the purity for the homogenization process
in the case with the ratio of characteristic times $T_1$ and $T_2$ taking the constant value $T_1/T_2=1/2$ while the
fixed point of the evolution varies from total mixture (unital process)
to a pure state (non-unital process). In particular, when the fixed point of the homogenization is a total mixture then
the homogenization dynamics in the C-P diagram is described by the line (1). This line lies below the line $C_{W}$
with the final point $C=0$ having the smallest value
of the purity. On the other hand, when the fixed point of the evolution is a pure state, then the corresponding C-P line (2) ends in the point
$C=0$ with the largest value of the purity $P=1/2$. }
\end{center}
\end{figure}
%%%%%%%%%%%%%%%%%%%%%%%%%%%%%%%%%%%%%

%%%%%%%%%%%%%%%%%%%%%%%%%%%%%%%%%%%
\begin{figure}
\label{cepe_homo_zmes}
\begin{center}
\includegraphics[width=7cm]{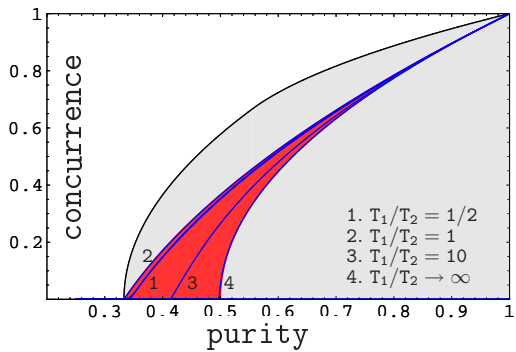}
\caption{(Color online)
The evolution of concurrence and purity for the homogenization
with the fixed point equal to the total mixture. In
this case the homogenization is a unital process. We vary the ratio
$T_1/T_2$ in the interval $[1/2,\infty]$.
For $T_1=T_2$ the homogenization exactly covers the line (2)
of Werner states ($C_{W}$).
In the limit of $T_1/T_2\to\infty$ this evolution coincides with the
line of decoherence ($C_{D}$).
}
\end{center}
\end{figure}
%%%%%%%%%%%%%%%%%%%%%%%%%%%%%%%%%%%%%

%%%%%%%%%%%%%%%%%%%%%%%%%%%%%%%%%%%
\begin{figure}
\label{cepe_homo_pure}
\begin{center}
\includegraphics[width=7cm]{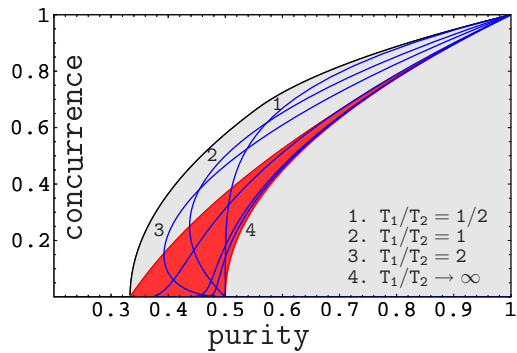}
\caption{(Color online)
The parametric plot
of the concurrence and the purity for the homogenization with the fixed point that corresponds
to a pure state. That is we consider an exponential decay into
a pure state. We vary the $T_1/T_2$ in the interval $[1/2,\infty]$.
In this case, the evolution ends in the point $C=0$ and $P=1/2$ irrespective of the particular
value of the ratio $T_1/T_2$. When $T_1/T_2=1/2$ the evolution is represented by the line which
is ``deep'' in the non-unital region of the C-P diagram. On the contrary, for $T_1/T_2\rightarrow\infty$
the corresponding line coincides with the bound $C_{D}$ (the decoherence line).
}
\end{center}
\end{figure}
%%%%%%%%%%%%%%%%%%%%%%%%%%%%%%%%%%%%%
%%%%%%%%%%%%%%%%%%%%%%%%%%%%%%%%%%%%%%%%%%%%%%%%%%%%%%%%%%%%%%%%%%%%%%%%%%%%%
\section{Conclusion}
%%%%%%%%%%%%%%%%%%%%%%%%%%%%%%%%%%%%%%%%%%%%%%%%%%%%%%%%%%%%%%%%%%%%%%%%%%%%%
In this paper we
have investigated how a local transformation (quantum channel described by a CP map)
of a sub-system affects
the entanglement and the global purity of the initial singlet state (or, equivalently,
arbitrary maximally entangled states) of two qubits.  We have analyzed the states
$\varrho_\cE=\cE\otimes\cI[\psi_s]$. We have used the concurrence and the purity of these states
to classify local channels $\cE$. In particular, using the concurrence-{\it vs.}-purity
diagrams we have specified the region induced by local unital channels. This region
does not cover the whole set of physically realizable states (see Fig.~2). Local unital channels
induce states that are represented by points in the region of the C-P diagram that is bounded
from below by the line $C_{D}$ and from the above the line $C_{W}$.
We have shown that even the most general (including non-unital) local channels acting just on a single qubit
of the original singlet state cannot generate MEMS. This means that the upper bound specified by the $C_{MEMS}$ line
cannot be achieved by the action of the local channel of the form $\varrho_\cE=\cE\otimes\cI[\psi_s]$. Specific achievable upper bound
for non-unital maps is to be determined. It definitely is above the line $C_W$ (except the values
of the concurrence $C=1$ and $C=0$ when it coincides with the bound on unital maps).
From our analysis it follows that for unital maps the lower bound of the achievable region is determined by the line
$C_D$. We conjecture, that this is also a general lower bound for non-unital maps. This conjecture is
supported by our numerical analysis, but we have no rigorous proof yet.

From our previous analysis an interesting observation follows. Specifically, if
the state $\varrho_\cE=\cE\otimes\cI[\psi_s]$ has a purity larger than 1/2 then the concurrence
has to be non-zero, correspondingly, the state is entangled. On the other hand, if the purity is less
than 1/3 then the state is separable.

In our paper we have analyzed neither the action of the bi-local channels ($\cE_1\otimes\cE_2$) nor
the action of the non-local maps $\cE_{12}$. It is clear, that using non-local maps any point of the C-P diagram
below the MEMS line $C_{\rm MEMS}$ can be achieved, i.e. the state $\varrho_\cE=\cE_{12}[\psi_s]$ can have arbitrary value
of the concurrence and the purity that is in the region specified by the bound $C_{\rm MEMS}$. On the other hand,
bi-local operations generate states $\varrho_\cE=\cE_1\otimes\cE_2[\psi_s]$ that are represented by the points
in a region of the C-P diagram that is restricted from above and from below. Specific boundaries are not known.
From particular examples one can conclude that if the two local transformations
are non-unital maps representing the exponential decay
then one can achieve states with zero concurrence but maximal purity.
We will analyze these bounds elsewhere.

We have studied in detail  time evolutions
described by Markovian semigroups.
In particular, we have shown that the decoherence process
is represented by a line that forms the lower bound $C_{D}$
of the achievable region for unital maps.
In our parametrization the time $t=0$ is represented by the
point $C=1$ and $P=1$, while the point of the entanglement destruction ($C=0$)
is achieved at some ``entanglement-breaking'' time $t_{sep}$
that is infinite - see discussion below.

The depolarization process saturates the upper
bound ($C_{W}$) of the region of maps induced
by unital maps. Here the entanglement-breaking time
is finite and equals $t_{sep}=T\ln 3$.
For dynamics that are described by non-unital maps
(e.g. homogenization processes) the entanglement-breaking
times can be both finite as well as infinite (as in the case
of the exponential decay).

We have paid attention only to evolutions governed by semigroups, i.e. by
Markovian processes. A general rule is that
for this type of time evolutions the associated C-P line
must be non-increasing, because local action cannot create entanglement. This property prevents
from loops in C-P diagram, but still allows that the purity
increases  while  the concurrence is decreasing
(see Fig.~6). For non-Markovian evolutions it is possible
to observe even loops, however, the question of more
general dynamics is out of the scope of this paper and will be discussed
elsewhere.

%%%%%%%%%%%%%%%%%%%%%%%%%%%%%%%%%%%%%%%%%%%%%%%%%%%%%%%%%%%%%%%%%%%%%%%%%%%%%
%%%%%%%%%%%%%%%%%%%%%%%%%%%%%%%%%%%%%%%%%%%%%%%%%%%%%%%%%%%%%%%%%%%%%%%%%%%%%
%%%%%%%%%%%%%%%%%%%%%%%%%%%%%%%%%%%%%%%%%%%%%%%%%%%%%%%%%%%%%%%%%%%%%%%%%%%%%
%%%%%%%%%%%%%%%%%%%%%%%%%%%%%%%%%%%%%%%%%%%%%%%%%%%%%%%%%%%%%%%%%%%%%%%%%%%%%

\noindent{\bf Acknowledgments}\newline
This work has been supported partially
by the European Union projects QGATES, CONQUEST, QUPRODIS,
and by Slovak government project APVT. We also acknowledge the
support of the Slovak Academy of Sciences via
the project CE-PI. VB would like to thank the Alexander von Humboldt Foundation for support.
We thank Carlos Pineda for interesting discussions.

%%%%%%%%%%%%%%%%%%%%%%%%%%%%%%%%%%%%%%%%%%%%%%%%%%%%%%%%%%%%%%%%%%%%%%%%%%%%%

%%%%%%%%%%%%%%%%%%%%%%%%%%%%%%%%%%%%%%%%%%%%%%%%%%%%%%%%%%%%%%%%%%%%%%%%%%%%%
%\end{multicols}


\begin{thebibliography}{00}

\bibitem{ckw}
V. Coffman, J. Kundu, and W.K. Wootters, {\it Distributed
entanglement}, {Phys. Rev. A} {\bf 61}, 052306 (2000)

\bibitem{koashi}
M. Koashi and A. Winter, {\it Monogamy of entanglement and other
correlations}, Phys. Rev. A {\bf 69}, 022309 (2004)

\bibitem{yu}
Ch. Yu and H. Song, {\it Multipartite entanglement measure}, Phys.
Rev. A {\bf 71}, 042331 (2005)

\bibitem{osborne}
T. Osborne and F. Verstraete, {\it General monogamy inequality for
bipartite qubit entanglement}, {\tt quant-ph/0502176}

\bibitem{wootters}
W.K. Wootters, {\it Entanglement of formation of an arbitrary
state of two qubits}, Phys. Rev. Lett. {\bf 80}, 2245 (1998)

\bibitem{ishizaka}
S.Ishizaka and T.Hiroshima, {\it Maximally entangled mixed states
in two qubits}, Phys. Rev. A {\bf 62}, 022310 (2000), {\tt
quant-ph/0003023}

\bibitem{verstraete}
F. Verstraete, K. Audenaert, T. De Bie, and B. De Moor, {\it
Maximally entangled states of two qubits}, Phys. Rev. A {\bf 64},
012316 (2001), {\tt quant-ph/0011110}

\bibitem{wei}
T.Ch. Wei, K. Nemoto, P.M. Goldbart, P.G. Kwiat, W.J. Munro, and
F. Verstraete, {\it Maximal entanglement versus entropy for mixed
quantum states}, Phys. Rev. A {\bf 67}, 022110 (2003)

\bibitem{adesso}
G. Adesso, F. Illuminati, and S. De Siena, {\it Characterizing
entanglement with global and marginal entropic measures}, Phys.
Rev. A {\bf 68}, 062318 (2003)

\bibitem{batle}
J. Batle, M. Casas, A. Plastino, and A.R. Plastino, {\it Maximally
entangled mixed states and concitional entropies}, Phys. Rev. A
{\bf 71}, 024301 (2005)

\bibitem{zyckowski}
K. Zyckowski, P. Horodecki, M. Horodecki, and R. Horodecki, {\it
Dynamics of quantum entanglement}, {\tt quant-ph/0008115}

\bibitem{eisert}
J. Eisert and M.B. Plenio, {\it A comparison of entanglement
measures}, J. Mod. Opt. {\bf 46}, 145 (1999)

\bibitem{grudka}
A. Miranowicz and A. Grudka, {\it Ordering two-qubit states with
concurrence and negativity}, Phys. Rev. A {\bf 70}, 032326 (2004)

\bibitem{nielsen}
M.A. Nielsen and  I.L. Chuang, {\it Quantum Computation and
Quantum Information}, (Cambridge University Press, Cambridge 2000)

\bibitem{presskill}
J. Preskill, {\it Quantum theory of Information and Computation},
available at\\ {\tt www.theory.caltech.edu/people/preskill}

\bibitem{ruskai}
M.B. Ruskai, S. Szarek, and E. Werner, {\it A characterizarion of
completely positive tracepreserving maps on
  ${\cal M}_2$},
Lin. Alg. Appl. {\bf 347}, 159 (2002)

\bibitem{ziman_deco}
M.Ziman and V.Bu\v zek, {\it All (qubit) decoherences: Complete
characterization and physical implementation}, to appear in
Phys.Rev.A {\bf 72}, to appear (2005), {\tt quant-ph/0505040}

\bibitem{ziman_homo}
M. Ziman, P. \v Stelmachovi\v c, V. Bu\v zek, M. Hillery, V.
Scarani, and N. Gisin, {\it Dilluting the quantum information},
Phys. Rev A {\bf 65}, 042105 (2002), {\tt quant-ph/0110164}

\bibitem{ziman_osid}
M. Ziman, P. \v Stelmachovi\v c, and V. Bu\v zek, {\it Description
of quantum dynamics of open systems based on collision-like
models}, Open systems and information dynamics {\bf 12}, 81 (2005)


\end{thebibliography}
\end{document}